\documentclass[reprint,amsmath,amssymb,floatfix]{revtex4-1}
\usepackage[utf8x]{inputenc}
\usepackage{graphicx}
\usepackage{dcolumn}
\usepackage{bm}
\usepackage{color}

\usepackage{soul} 
\usepackage[normalem]{ulem} 

\begin{document}

\preprint{APS/123-QED}

\title{Scaling and renormalization in the modern theory of
  polarization: application to disordered systems}

\author{Bal\'azs Het\'enyi$^{1,2,3}$, Sel\c{c}uk Parlak$^3$, and Mohammad Yahyavi$^3$}
\affiliation{$^1$MTA-BME Topology and Correlations Research Group, Department of Physics, Budapest
  University of Technology and Economics, H-1521 Budapest, Hungary \\ $^2$Department of Theoretical Physics, Budapest University of Technology and Economics, Budafoki út 8, H-1111, Budapest, Hungary
  \\ $^3$ Department of Physics, Bilkent University, TR-06800
  Bilkent, Ankara, Turkey }

\date{\today}

\begin{abstract}
We develop a scaling theory and a renormalization technique in the
context of the modern theory of polarization.  The central idea is to
use the characteristic function (also known as the polarization
amplitude) in place of the free energy in the scaling theory and in
place of the Boltzmann probability in a position-space renormalization
scheme.  We derive a scaling relation between critical exponents which
we test in a variety of models in one and two dimensions. We then
apply the renormalization to disordered systems.  In one dimension the
renormalized disorder strength tends to infinity indicating the entire
absence of extended states.  Zero(infinite) disorder is a
repulsive(attractive) fixed point.  In two and three dimensions, at
small system sizes, two additional fixed points appear, both at finite
disorder, $W_a$($W_r$) is attractive(repulsive) such that $W_a<W_r$.
In three dimensions $W_a$ tends to zero, $W_r$ remains finite,
indicating metal-insulator transition at finite disorder.  In two
dimensions we are limited by system size, but we find that both $W_a$
and $W_r$ decrease significantly as system size is increased.
\end{abstract}


\maketitle

\section{Introduction}
\label{sec:intro}

Scaling theory of localization in disordered
systems~\cite{Abrahams79,Lee85,Evers08,Langedijk09} has a long
history.  A milestone work by Abrahams {\it et al.}~\cite{Abrahams79},
often referred to as the ``gang of four'' paper (G4), put forth this
theory to explain the dimensional dependence of criticality.  The
central results for systems of no symmetry are that all states are
localized in one dimension (1D), even in two dimensions (2D) there are
no extended states, but here a crossover occurs, and only in three
dimensions (3D) does a true metal-insulator transition occur in the
form of an unstable fixed point.

Experimental evidence shows unambiguous support for the G4 conclusions
in 1D~\cite{Billy08,Roati08} and
3D~\cite{Kondov11,Jendrzejewski11,Semeghini15}.  The 2D results were
more difficult to establish~\cite{White20} experimentally, due to the
possibility of weak
localization~\cite{RobertdeSaintVincent10,Jendrzejewski12,Muller15}.
Theoretically a debate~\cite{Kuzovkov02,Markos04,Suslov05} about a
possible metal-insulator transition in 2D, rather than the entire
absence of extended states, arose.

An important development in the understanding of crystalline systems
in general (with or without disorder) was the development of the
modern theory of polarization~\cite{King-Smith93,Resta94,Resta00}
(MTP).  This theory provides the tools~\cite{Resta98,Resta99} to
measure localization.  Although some
studies~\cite{Bendazzoli10,Varma15,Olsen17} have used these tools to
assess localization in disordered systems, how the G4 results concur
with the MTP is still an open question.

In this paper we seek to fill this gap by developing thermodynamic
scaling and renormalization
methods~\cite{Reichl16,Kadanoff66,Wilson71} within the MTP context.
The idea is to use the MTP characteristic function (also known as the
polarization amplitude~\cite{Kobayashi18}) in place of the partition
function as a starting point for both scaling (similar to Widom
scaling of critical exponents) and our position space renormalization
scheme.

In 1D our flow lines tend to a high disorder attractive fixed point,
meaning that there are no extended states.  In 2D and 3D, for finite
system sizes, we find an attractive ($W_a$) and a repulsive ($W_r$)
fixed point ($W_a<W_r$).  The key question becomes how these fixed
points evolve with system size.  In 3D $W_a$ tends to zero, while
$W_r$ tends to a finite number, indicating a metal-insulator
transition.  In 2D both the $W_a$ and $W_r$ decrease, but due to
finite size limitations it is difficult to draw a definite conclusion
valid for the thermodynamic limit.  We discuss the possible scenarios.

Other renormalization approaches to Anderson localized systems, apart
from the G4 scaling theory, also have been
used~\cite{Domany79,Sarker81,Shapiro82,Foster09}.  Traditional
real-space renormalization schemes, based on blocking sites of the
lattice (Migdal-Kadanoff procedure) concur with G4.  In the context of
MTP renormalization has been applied by Voit and
Nakamura~\cite{Nakamura02}, but this technique relies on bosonization
and is only applicable in 1D.

There are many
investigations~\cite{Marzari12,Yahyavi17,Kobayashi18,Hetenyi19,Patankar18,Furuya19}
which focus on the distribution function and the scaling of the
polarization amplitude in band theoretic and correlated quantum
models.  MTP also serves as the starting point for deriving
topological invariants~\cite{Bernevig13}, for example, the
time-reversal polarization, the topological invariant in the Fu-Kane
spin-pump~\cite{Fu06} is the difference of the Zak phases of different
members of a Kramers pair.

In section \ref{sec:bckgrnd} the scaling theory of localization and
the MTP are outlined, and the motivation of this work is stated, also
placing our work in contemporary context.  In section \ref{sec:models}
the Hamiltonians of the models studied in this work are given.  In
section \ref{sec:Widom} we derive a relation between critical
exponents based on Widom's thermodynamic scaling and test it for some
model systems.  In section \ref{sec:renorm} we develop our
renormalization approach and apply it to disordered systems in
different dimensions.  In section \ref{sec:cnclsns} we conclude our
work.
 
\section{Background and Motivation}
\label{sec:bckgrnd}

The starting point of the G4 scaling theory~\cite{Abrahams79} is the
specification of the Thouless number~\cite{Edwards72} as the relevant
quantity to analyze.  The Thouless number is a dimensionless
conductance defined as
\begin{equation}
  g(L) = \frac{G}{e^2/2\hbar}, 
\end{equation}
where 
\begin{equation}
  G = \frac{\Delta E}{dE/dN},
\end{equation}
where $\Delta E$ is the difference between energy levels calculated
using periodic boundary conditions and anti-periodic boundary
conditions, $dE/dN$ is the average level spacing.  The argument in
support of $G$ as a conductance is that it is localization that
determines whether a state is insulating or not.  A delocalized state
should be sensitive to changing the boundary conditions, whereas a
localized one should not.  G4 then argues that the $g(L)$ depends only
on the system size.  The scaling theory then analyzes the scaling
function
\begin{equation}
  \label{eqn:dlngdlnL}
  \beta(g) = \frac{d \ln g}{d \ln L}.
\end{equation}
In this function $L$ appears explicitly, since in disordered systems
the system size dependence is more pronounced than in clean systems.
Asymptotic analysis can be applied.  When the conductance is small
(states are localized), it is expected that
\begin{equation}
  \label{eqn:gexpL}
  g = \exp(-L/\xi),
\end{equation}
where $\xi$ denotes a correlation length.  When extended states
dominate, the conductance is expected to behave as
\begin{equation}
  \label{eqn:gsigma}
  g = \sigma_0 L^{d-2},
\end{equation}
where $\sigma_0$ denotes the conductivity, and $d$ is the
dimensionality of the system.  From this information, by plotting
$\beta(g)$ as a function of $\ln g$ the dimensional dependence of the
critical behavior can be surmised.  In 1D, all states are localized,
since the function $\beta(g)$ is always negative, even as $\ln g$ goes
to infinity.  In 2D the curve is negative when $\ln g$ is negative,
and it approaches zero as $\ln g$ goes to infinity, meaning that even
in 2D the states are extended, however, in this case due to the
crossover between exponential and logarithmic behavior, G4 predicts
that experiments may detect a sharp mobility edge.  In 3D, since the
curve crosses $\beta(g) = 0$, corresponding to an unstable fixed
point, a metal-insulator transition is predicted.

Overall, in assessing a metal-insulator transition (in ordered or
disordered) many-body quantum systems, the 1964 work of
Kohn~\cite{Kohn64} was a starting point.  On the one hand, Kohn
argued~\cite{Kohn64} that assessing whether a system is metallic or
insulating can be done by investigating the response of the system to
twisting the boundary conditions.  On the other hand, Kohn also
pointed out that the localization of the center of mass of the charge
distribution is the ultimate measure of whether a system is conducting
or insulating.  The two approaches are equivalent.

The central difficulty addressed by MTP was the ill-defined nature of
the position operator in systems with periodic boundary conditions.
This difficulty hindered the application of the hypothesis of
Kohn~\cite{Kohn64} in calculations for crystalline systems.  The
problem was overcome by casting the polarization in terms of a
geometric phase~\cite{Berry84} of the Zak~\cite{Zak89} variety, which
arises upon integrating across the Brillouin zone.  Other relevant
properties, such as the many-body generalization~\cite{Resta98} of the
polarization, and the variance thereof~\cite{Resta99} were also
derived.  The program of Kohn~\cite{Kohn64} was later realized through
the MTP~\cite{King-Smith93,Resta94,Resta00,Resta99}.

In the formalism of Resta and Sorella, the variance of the total
position is cast~\cite{Resta99} as
\begin{equation}
  \label{eqn:sigma}
  \chi^{(2)} = \sigma^2 = -\frac{L^2}{2 \pi^2} \mbox{Re} \ln Z_1
\end{equation}
where
\begin{equation}
  \label{eqn:Z_q}
  Z_q = \langle \Psi_0 |\exp \left( i \frac{2 \pi q}{L} \hat{X}
  \right) |\Psi_0 \rangle,
\end{equation}
and where $|\Psi_0 \rangle$ denotes a quantum ground state, $q$ is an
integer, and the total position operator is defined as
\begin{equation}
  \label{eqn:hatX}
  \hat{X} = \sum_{j=1}^L \hat{n}_j j,
\end{equation}
where $\hat{n}_j$ is the density operator at site $j$.  If $\sigma$
tends to infinity with system size, the system is metallic.

In this work we perform calculations for disordered systems in
different dimensions based on MTP.  Since the Thouless number is a
measure of localization, we replace it with the the quantity from the
MTP which can be taken as its analog,
\begin{equation}
  g(L) \rightarrow 1 - |Z_1|.
\end{equation}
This is not an exact correspondence by any means.  We justify it by
first stating that the variance can be cast according to an
approximation~\cite{Hetenyi19} different from the one of Resta and
Sorella,
\begin{equation}
  \label{eqn:sig_HD}
  \sigma^2 = \frac{L^2}{2 \pi^2} (1 - |Z_1|).
\end{equation}
This approximation has the advantage that in the limit of the Fermi
sea ($Z_1 = 0$) $\sigma$ scales with system size linearly, which is
expected on robust physical grounds.  Furthermore, in
Fig. \ref{fig:flowlines}(a), in the inset, we show the quantity
$1-|Z_1|$ as a function of the Thouless number (defined as the sum
over the absolute value of the difference in energy between periodic
and anti-periodic boundary conditions, divided by the total energy
difference), for a 1D system of $L=160$, and averaged over one hundred
disorder realizations (replicas) (Hamiltonian in
Eq. (\ref{eqn:H_disorder})).  The function is monotonic, which also
justifies our replacing of the Thouless number with $|Z_1|$ in our
renormalization scheme (section \ref{sec:renorm}).

To perform the asymptotic analysis, the G4 scaling theory uses two
pieces of information. In the small conductivity (large disorder)
limit it is assumed that the conductivity localizes exponentially
(Eq. (\ref{eqn:gexpL})).  In the opposite limit the conductance is
related to the conductivity via the relation (Eq. (\ref{eqn:gsigma})).
In the latter, there appears to be no direct prescription to calculate
the conductance based on a microscopic Hamiltonian.  To phrase the
question differently: given a disordered Hamiltonian for which we can
calculate the eigenstates, how do we calculate the conductance?  Which
states do we consider?  Should we consider a distribution of states?
In our calculations below we will average the relevant quantities over
all states.  This amounts to a high-temperature approximation, since
all states have the same contribution.  The position operator we
use~\cite{Selloni87,Fois88,Ancilotto92} is a single-body one.

\section{Models}
\label{sec:models}
 
Most of this paper is devoted to disordered systems.  The one-dimensional
version of the disordered Hamiltonian we study can be written
\begin{equation}
  \label{eqn:H_disorder}
  \hat{H} = \sum_{i=1}^L \left[-t(\hat{c}_i^\dagger \hat{c}_{i+1} + \mbox{H.c.}) + W \xi_i \hat{n}_i \right],
\end{equation}
where $t$ denotes the hopping parameter (which will be taken as the
unit of energy), $W$ indicates the disorder strength, and $\xi_i$ is a
normal distributed random number.  On the other hand, in the next
section we also test the result of our Widom scaling theory for the
Su-Schrieffer-Heeger~\cite{Su79} (SSH) and Rice-Mele~\cite{Rice82}
(RM) models.  The latter can be written,
\begin{equation}
  \label{eqn:H_SSH_RM}
  \hat{H} = \sum_{i=1}^L \left[-(t + (-1)^{i}\delta t)(c_i^\dagger
    c_{i+1} + \mbox{H.c.}) + \Delta (-1)^i \hat{n}_i \right],
\end{equation}
where $\delta t$ denotes the alternation in hopping, and $\Delta$
denotes the strength of an alternating on-site potential.  The SSH
Hamiltonian is obtained by setting $\Delta=0$ in
Eq. (\ref{eqn:H_SSH_RM}).

\section{Widom scaling in the modern theory of polarization}
\label{sec:Widom}

We consider an $N$-electron system, one-dimensional for convenience,
periodic in $L$.  The discrete analog of the characteristic function
is given in Eqs. (\ref{eqn:Z_q}) and (\ref{eqn:hatX}).  The expression
for the average position in a many-body crystalline system given by
Resta is
\begin{equation}
  \label{eqn:Pol}
  \langle X \rangle = \frac{L}{2\pi} \mbox{Im} \ln Z_1,
\end{equation}
which reduces to the Zak phase if $\Psi_0$ is a Slater determinant.
The average position~\cite{Resta98} (Eq. (\ref{eqn:Pol})) and the
variance~\cite{Resta99} (Eq. (\ref{eqn:sigma})) are a finite
difference derivatives of $\ln Z_q$ at $q=0$, first and second
derivatives, respectively.

To our purposes in this section it is more suitable to take the
thermodynamic limit in Eq. (\ref{eqn:Z_q}) and write
\begin{equation}
  \label{eqn:Z_K}
  Z(K) = \langle \Psi_0 |\exp \left( i K \hat{X} \right) |\Psi_0
  \rangle,
\end{equation}
and express the variance as
\begin{equation}
  \label{eqn:chi2}  
  \chi^{(2)} = - \left. \frac{\partial^2 \ln Z(K)}{\partial K^2}
  \right|_{K=0}.
\end{equation}
The quantity $\chi^{(2)}$ can be interpreted as a semi-classical
approximation to the dielectric susceptibility.  It is to be expected
that if a phase transition point is approached from the insulating
side $\chi^{(2)}$ diverges.

To keep the discussion general we introduce a variable $D$ which
characterizes the approach to the critical point (we assume that $D=0$
is a quantum phase transition, or gap closure point).  We first define
\begin{equation}
  \label{eqn:chi_n}
  \chi^{(n)}(D,K) = \frac{1}{i^n} \frac{\partial^n \ln
    Z(D,K)}{\partial K^n},
\end{equation}
and also define three critical exponents characterizing the approach
to the phase transition:
\begin{eqnarray}
  \label{eqn:chi_bdg}
  \chi^{(2)}(D,0) &\propto& \frac{1}{D^\beta}, \\ \nonumber
  \chi^{(4)}(D,0) &\propto& \frac{1}{D^\gamma}, \\
  \chi^{(2)}(0,K) & \propto & \frac{1}{K^\delta}. \nonumber
\end{eqnarray}
We also define the singular quantity,
\begin{equation}
  \Phi(D,K) = -\ln Z(D,K),
\end{equation}
which serves as the analog of the free energy.  Applying the scaling
relation $ \Phi(\lambda^a D,\lambda^b K) = \lambda
\Phi(D,K)$, we obtain the following relation between the scaling
exponents:
\begin{equation}
  \label{eqn:bdg}
  \gamma \delta = \beta (\delta + 2).
\end{equation}
\begin{figure}[ht]
 \centering
 \includegraphics[width=7cm,keepaspectratio=true]{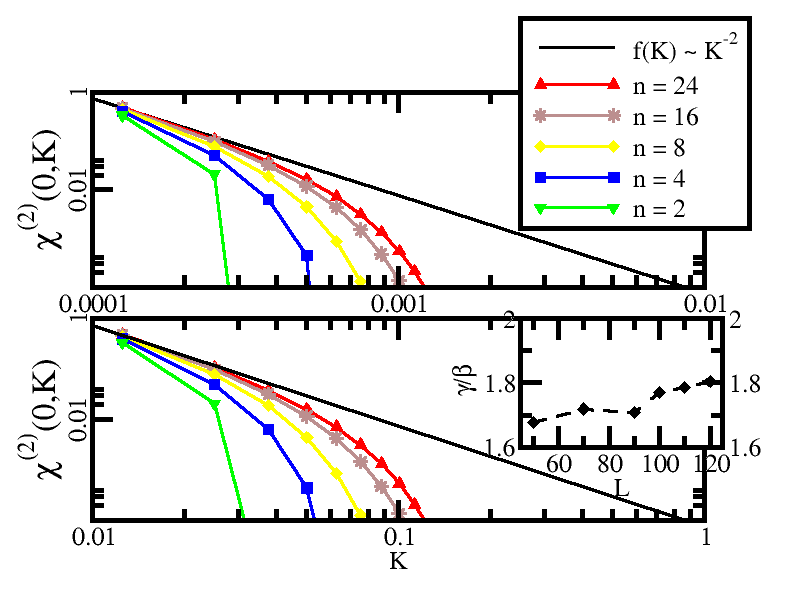}
 \caption{Log-log plot of $\chi^{(2)}(0,K)$ as a function of $K$ for
   different order finite difference approximations to the second
   derivative (see Eq. (\ref{eqn:chi_n})) for two system sizes.  $n$
   refers to the order of the approximation, $\mathcal{O}( K^{-2n})$.
   The upper panel shows results for $L=100000$, the lower panel for
   $L=1000$.  The solid line indicates a curve of the form $f(K)
   \propto K^{-2}$.  The legend is for the upper and lower panels, not
   the inset of the lower panel.  The inset of the lower panel shows
   the estimated ratio of the exponents $\gamma$ and $\beta$ as a
   function of system size for a two-dimensional disordered system.}
 \label{fig:chi2_K}
\end{figure}

We tested this relation on a number of different models.  For the
exponent $\delta$, ($D = 0$ limit corresponds to the Fermi sea)
our calculations are shown in Fig. \ref{fig:chi2_K}.  The different
panels show different system sizes, the different curves within each
panel are finite difference approximations to the second derivative
(Eq. (\ref{eqn:chi_bdg})).  From these calculations we can conclude
that $\delta=2$, which also means that $\gamma = 2 \beta$.  We then
calculated the other exponents of the SSH and Rice-Mele
(Eq. (\ref{eqn:H_SSH_RM})) models, in 1D and 2D, and found that
$\beta=2$ and $\gamma=4$.

In addition, we also calculated the exponents for a 1D disordered
system with the Hamiltonian given by Eq. (\ref{eqn:H_disorder}).  When
calculating the cumulants (Eq. (\ref{eqn:chi_bdg})) a difficulty faced
was that a smooth curve only results for large systems.  In
particular, at small $W$, large system sizes are needed for converged
results.  We calculated the critical exponents for an $L=1280$ system,
averaged over one hundred realizations of the disorder and found
$\beta = 3.61(5)$ and $\gamma=7.16(27)$, consistent with
Eq. (\ref{eqn:bdg}).  We also made calculations for three different
realizations of the disorder (three different calculations no disorder
average is taken, only $W$ was swept) for a system of size $L=2560$.
The results for the exponents turned out to be $\beta =
3.53(4),3.45(4),3.52(5)$ and $\gamma = 7.12(9),6.90(12),7.07(11)$,
respectively, again consistent with Eq. (\ref{eqn:bdg}) within error
bars.

\begin{figure}[ht]
 \centering 
 \includegraphics[width=6.5cm,keepaspectratio=true]{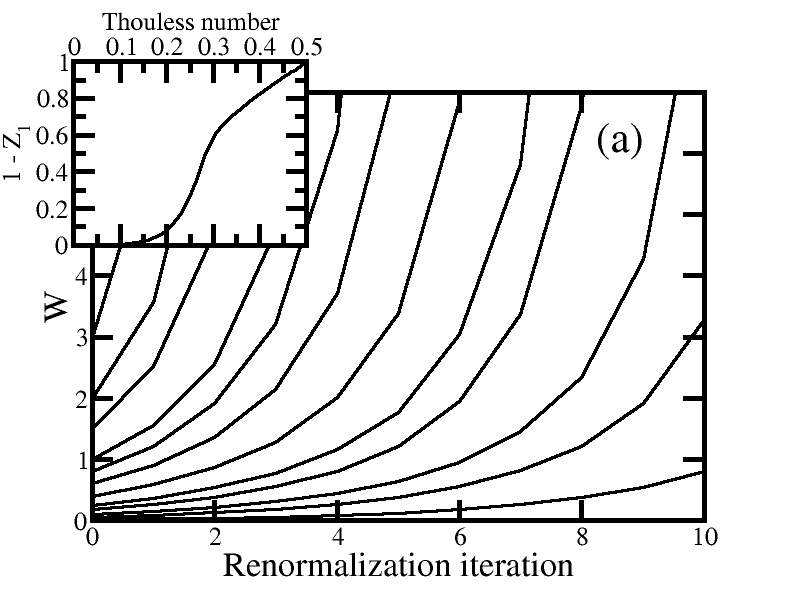}
 \includegraphics[width=6.5cm,keepaspectratio=true]{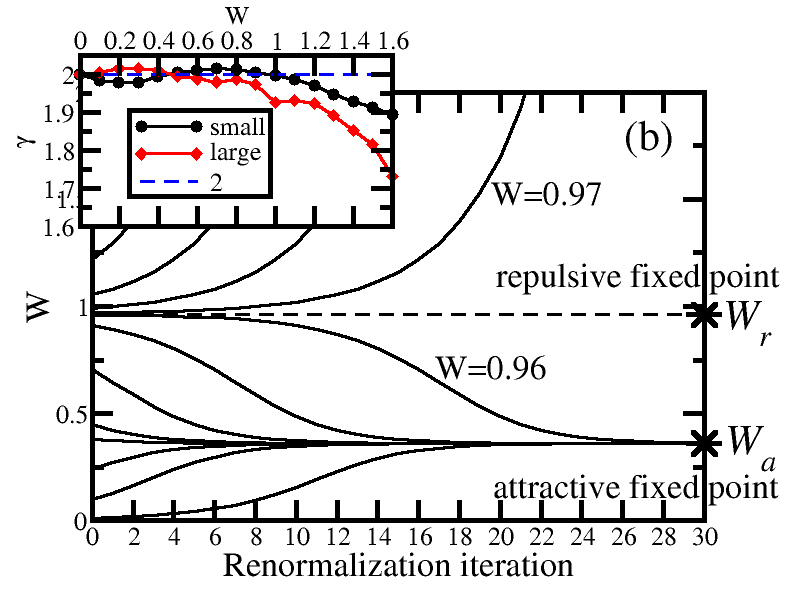}
 \includegraphics[width=6.5cm,keepaspectratio=true]{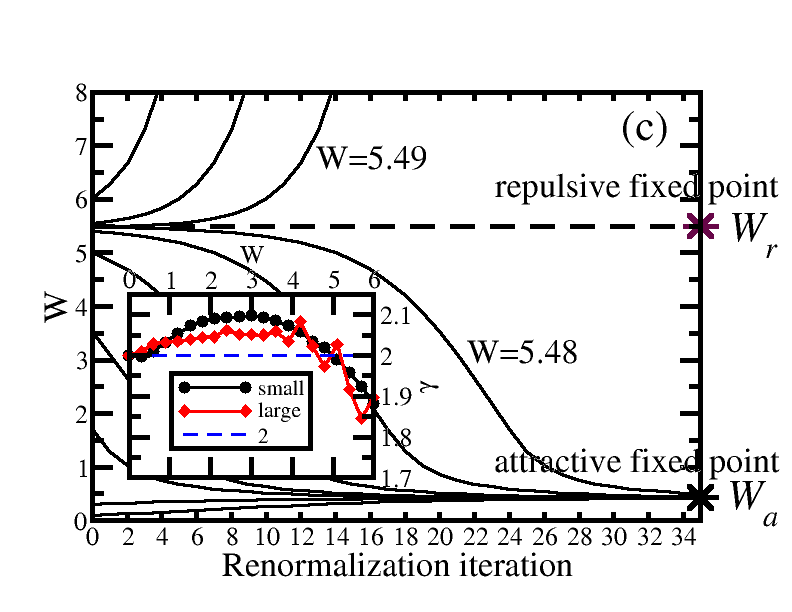}
 \caption{Renormalization flow lines for disordered systems of
   different dimensions.  (a) One-dimensional $L=160$, 100 replicas
   averaged.  The inset shows the quantity $1-Z_1$ vs. the Thouless
   number.  (b) Two-dimensional, $L=24$, 100 replicas averaged.  The
   repulsive ($W_r$) and attractive ($W_a$) fixed points are also
   indicated.  The inset shows the size scaling exponent calculated
   using system sizes of $L=12,14,16,18$ (small) and $L=72,80,96$
   (large).  (c) Three-dimensional, $L=8$, 100 replicas averaged.
   The repulsive ($W_r$) and attractive ($W_a$) fixed points are also
   indicated.  The inset shows the size scaling exponent calculated
   using system sizes of $L=6,8,10$ (small) and $L=20,22,24$ (large).
 }
 \label{fig:flowlines}
\end{figure}

We also performed a calculation for a 2D disordered system.  In this
calculation we encountered system size related difficulties, however
we provide an estimate of the ratios of critical exponents (inset of
the lower panel of Fig. \ref{fig:chi2_K}).  As shown below, at the
system sizes accessible, there is a transition in two-dimensional
disordered systems (see Fig. \ref{fig:flowlines}), but the position of
this transition decreases as system size is increased (see
Fig. \ref{fig:Wr23D}, upper panel).  The relevant range of
$\chi^{(2)}$ and $\chi^{(4)}$ as a function of $W$ is above the
transition (which itself changes with system size), but before the
disorder becomes too large, because then the errors are larger than
the values of $\chi^{(2)}$ and $\chi^{(4)}$.  We estimated the
critical exponents $\gamma$ and $\beta$ by looking for the maxima of
the derivatives of the functions $\log \chi^{(2)}$ and $\log
\chi^{(4)}$ as a function of the logarithm of disorder strength.  This
estimate is shown in the inset of the lower panel of
Fig. \ref{fig:chi2_K}.  The estimated ratio is increasing of the
range of system sizes studied, and it is $\gamma/\beta \approx 1.8$
for the largest size, $L=120$.

\section{Real-space renormalization in the modern theory of polarization}
\label{sec:renorm}

In real-space renormalization~\cite{Reichl16,Kadanoff66,Wilson71} one
starts with the blocking of sites of the lattice.  To each block a
single block variable is assigned.  The blocked system is assumed to
have the same Hamiltonian as the original Hamiltonian (although, this
can mean an extended set of couplings).  The parameters of the blocked
Hamiltonian are tuned to produce the same Boltzmann probability as the
original Hamiltonian, provided that the configurations of the starting
Hamiltonian consistent with a given configuration of the block
variables are summed over (traced out).

A common way to define a blocked system is by the procedure of
``decimation'', in which some of the variables of the original system
are traced out.  In most cases the set of equations obtained this way
are overdetermined, so either new parameters have to be introduced
(for example, by extending the couplings included in the Hamiltonian)
or via introducing further approximations, for example equating
cumulants~\cite{Niemeijer73}, between the original and the
renormalized system, rather than the full Boltzmann distribution.

We now apply this set of steps to the quantity
\begin{equation}
 \tilde{P}_j(x_1,...,x_L) = |\Psi_0(x_1,...,x_L)|^2\exp\left(i \frac{2
   \pi}{L} \hat{X}\right).
\end{equation}
Here, the index $j$ indicates the step in the renormalization.  As a
first step we integrate out all the odd sites, resulting in
\begin{equation}
 \tilde{P}_{j+1}(x_2,...,x_L) = \int.\int d x_1 ... d x_{L-1} \tilde{P}_j(x_1,...,x_L).
\end{equation}
We now require that the $\tilde{P}_j(x_2,...,x_L)$ of the remaining
variables equals $\tilde{P}_{j+1}(x_2,...,x_L)$.  It is easy to see,
based on the definition of $\hat{X}$ (Eq. (\ref{eqn:hatX})), that
$\tilde{P}_j(x_2,...,x_L)$ corresponds to the distribution of a system
with half the size of the original one, $L/2$.  As in other real-space
renormalization techniques, the requirement is too stringent, so to
arrive at a practical scheme, we use the relaxed requirement,
\begin{equation}
  Z_1(W_{j+1},L/2) = Z_1(W_j,L).
\end{equation}
In other words, in each renormalization step, with a given disorder
strength and system size, we find the disorder strength at half the
system size which generates the same value of $Z_1(W_j,L)$.  It is not
the entire probability distribution that is kept fixed in the course
of a renormalization step, but only one Fourier mode of the
distribution of the many-body position.  In this sense, this
renormalization is taylored to MTP.

\begin{figure}[ht]
 \centering
 \includegraphics[width=7cm,keepaspectratio=true]{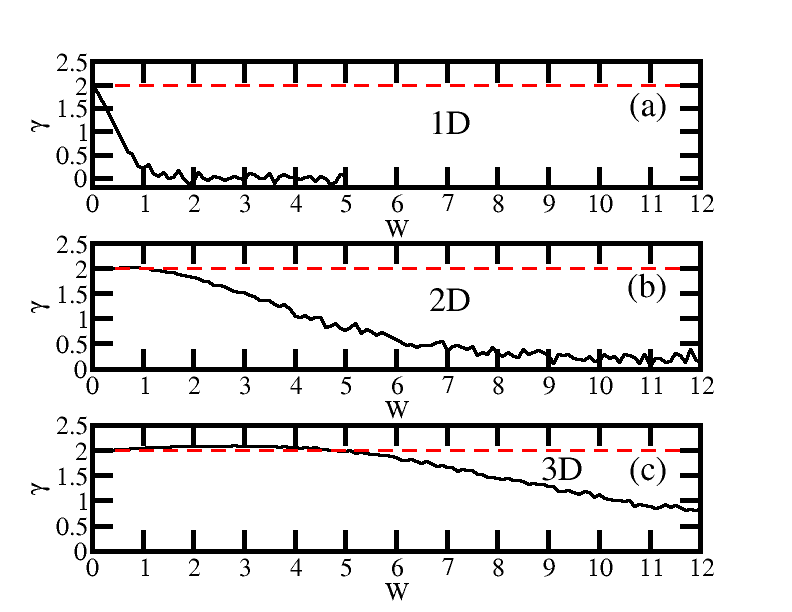}
 \caption{Size scaling exponent of the variance of the polarization
   for a (a) one-dimensional, (b) two-dimensional, and (c)
   three-dimensional system.}
 \label{fig:gam_d}
\end{figure}

In Fig. \ref{fig:flowlines} the flowlines of the renormalization
scheme are shown for systems of different dimensions.
Fig. \ref{fig:gam_d} shows the size scaling exponent ($\gamma$) of the
variance of the polarization.  We define $\gamma$ as
\begin{equation}
  \chi^{(2)} = a L^\gamma.
\end{equation}
In previous studies~\cite{Hetenyi19} it was shown that metal-insulator
transitions can be accurately determined by investigating $\gamma$.
In clean systems a gapless system will exhibit $\gamma=2$, while in
gapped insulators $\gamma \rightarrow 1$.

 Fig. \ref{fig:flowlines}(a) shows results for a 1D disordered system
 of size $L=160$, and one hundred realizations of the disorder
 averaged. All flow lines which start at a finite value tend towards
 infinity indicating that all states are localized.  We found no
 significant size dependence or dependence on the number of replicas.
 Also, the upper panel of Fig. \ref{fig:gam_d}(a) is consistent with
 the renormalization flows.  The clean conducting system ($W=0$)
 exhibits $\gamma=2$, and finite disorder strength leads to a rapid
 decrease in $\gamma$.  These results concur with the
 G4~\cite{Abrahams79}.

Fig. \ref{fig:flowlines}(b) shows the flowlines in a 2D sample
calculation.  The linear dimension of the system is $L=24$, the
calculation was done on a square lattice.  The number of disorder
configurations averaged was one hundred.  The flow lines here show
qualitatively different behavior from the 1D case.  We find two fixed
points on the $W$ axis, one repulsive, $W_r\approx1$, above which the
flow lines tend to infinity, corresponding to a fully localized state.
Below $W_r$, the flow lines which start at finite disorder strength
tend to a finite disorder strength of $W_a$.  $W_r$($W_a$) is a
repulsive(attractive) fixed point.  We have done a number of
calculations and this qualitative behavior is maintained, however, we
found variation in the values of $W_r$ and $W_a$.
Fig. \ref{fig:gam_d}(b) shows the size scaling exponent in 2D.  In
these calculations system sizes up to $L=18$ ($L$ is the linear
dimension) were used, and one hundred replicas were averaged.  Note
that until $W \approx 1$, the size scaling exponent is approximately
two, above that value it decreases.  The flow lines in
\ref{fig:flowlines}(b) are consistent with the behavior of the scaling
exponent.

Similar behavior is found in 3D, (see Figs. \ref{fig:flowlines}(c) and
\ref{fig:gam_d}(c)).  A repulsive fixed point $W_r$ and an attractive
fixed point $W_a$ and $W_r>W_a$.  The flow lines shown in
Fig. \ref{fig:flowlines}(c) are for an $8\times 8 \times 8$ system
with 100 replicas averaged.

\begin{figure}[ht]
 \centering
 \includegraphics[width=7cm,keepaspectratio=true]{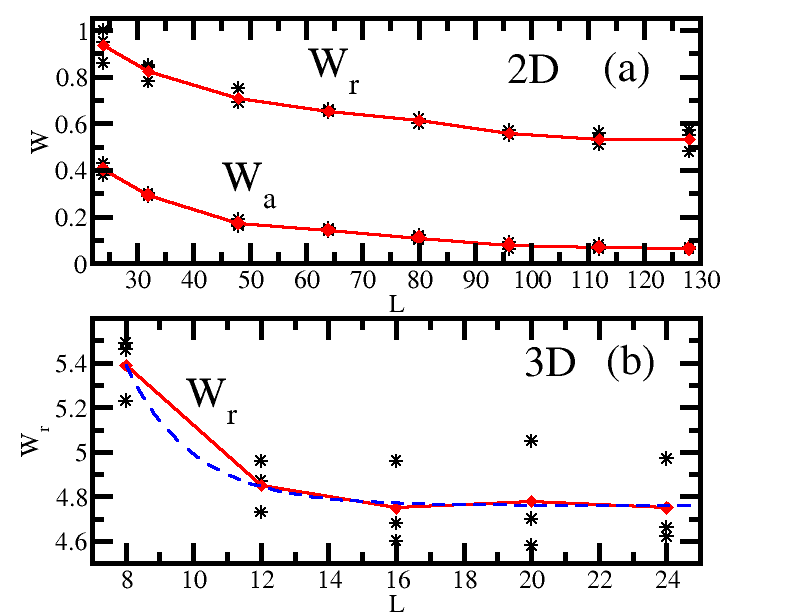}
 \caption{The behavior of fixed points as a function of system size in
   2D and 3D.  In 2D both the repulsive and the attractive fixed
   points are shown.  The asterisks correspond to the raw data (three
   calculations) also shown in Tables \ref{tab:2D} and \ref{tab:3D}.
   The red diamonds show the average of the three calculations for
   each system size.  The blue dashed curve for 3D shows a fit
   function (see the text for a detailed discussion).}
 \label{fig:Wr23D}
\end{figure}

In the G4 scaling theory the system size itself appears as a relevant
variable (Eq. (\ref{eqn:dlngdlnL})).  For this reason, we further investigate
the system size dependence of the attractive and repulsive fixed
points.  Our results are shown in Tables \ref{tab:2D} and \ref{tab:3D}
and Fig. \ref{fig:Wr23D}).  To keep the CPU time manageable, we
reduced the number of disorder configurations averaged,
proportionately to system size.  We made three calculations for each
data point, and the raw data is shown in Tables \ref{tab:2D} and
\ref{tab:3D}.  A plot is shown in Fig. \ref{fig:Wr23D}, with the raw
results and their average.

In 3D $W_a$ tends to zero, and $W_r$ approaches a finite value with
large system size.  The function fit in the lower panel of
Fig. \ref{fig:Wr23D} is $f(L) = a \exp(b L) + c$, where $a=31.8259$,
$b=-0.489526$ and $c=4.75641$.  A metal-insulator transition takes
place at finite $W_r$, in agreement with the G4~\cite{Abrahams79}.  In
2D, due to finite size limitations, a definite conclusion is difficult
to reach.  Several scenarios are possible, depending on what happens
to $W_r$ and $W_a$ as $L$ becomes large.  Our results (upper panel,
Fig. \ref{fig:Wr23D}) show a sizeable decrease for both $W_r$ and
$W_a$, $W_r \approx 1$ for the smallest size, and $W_r \approx 0.53$
for the largest one.  Also, while $W_a$ does decrease, we can not say
that it reaches zero (as it happens in 3D).  There are then several
possible scenarios.  If $W_a$ and $W_r$ both go to zero, then extended
states are absent, concurring with the G4.  If only $W_a$ goes to
zero, that would correspond to a metal-insulator transition, which
would appear to be in accordance with transfer matrix and Lyapunov
exponent based calculations~\cite{Kuzovkov02,Suslov05}.  A third
possibility is that $W_a$ and $W_r$ both tend to finite values in the
thermodynamic limit such that $W_r>W_a$.  This would mean that there
is a transition, but the small disorder states would be localized,
rather than extended, even though, they would scale with system size
as $L$ (substitute a finite size dependent $Z_1$ in
Eq. (\ref{eqn:sig_HD})).  We note that localized states of a
qualitatively different nature, exhibiting power-law decay rather than
exponential, have also been suggested~\cite{Mott85} in 2D.

We also calculated $\gamma$ using different sets of systems, the
results shown as insets in Fig. \ref{fig:flowlines} middle and lower
panel.  In 2D we used systems with linear extension
$L=12,16,18$($L=72,80,96$) for the data points designated
``small''(``large'').  For the former, $\gamma$ remains two up to $W
\approx 1.0$, while for the latter, the decrease from two starts
earlier at $W \approx 0.6$.  In 3D increasing the system size does not
lead to a similar decrease.  (Here the linear extensions were
$L=6,8,10$($L=20,22,24$) for ``small''(``large'').)  Although, these
results are also limited by system size limitations, the suggestion is
that when the system size is increased the curve in
Fig. \ref{fig:gam_d}b eventually becomes like Fig. \ref{fig:gam_d}a,
in other words, no sign of extended states remains, and our results
likely concur with G4.
\begin{table}[t]
\begin{center}
  \begin{tabular}{ |c||c|c|c||c|c|c| }
 \hline
 Size(No. of replicas) & $W_r(1)$ & $W_r(2)$ & $W_r(3)$ & $W_a(1)$ & $W_a(2)$ & $W_a(3)$  \\ 
 \hline
 $24\times 24$($100$) & $1.00$ & $0.86$& $0.95$ & $0.38$ & $0.40$ & $0.43$ \\ 
 $32\times 32$($50$) & $0.78$ & $0.85$& $0.84$ & $0.30$ & $0.29$ & $0.29$ \\ 
 $48\times 48$($25$) & $0.69$ & $0.75$& $0.69$ & $0.19$ & $0.17$& $0.16$ \\ 
 $64\times 64$($15$) & $0.65$ & $0.66$& $0.65$  & $0.15$ & $0.14$ & $0.14$ \\ 
 $80\times 80$($10$) & $0.60$ & $0.62$ & $0.62$  & $0.11$ & $0.10$ &  $0.12$ \\ 
 $96\times 96$($5$) & $0.57$ & $0.55$ & $0.55$ & $0.06$ & $0.09$ &  $0.09$ \\ 
 $112\times 112$($4$) & $0.51$ & $0.53$ & $0.56$ & $0.08$ & $0.07$ &  $0.06$ \\ 
 $128\times 128$($3$) & $0.48$ & $0.55$ & $0.57$ & $0.06$ & $0.07$ &  $0.06$ \\ 
 \hline
\end{tabular}
\end{center}
    \caption{The repulsive ($W_r$) and attractive ($W_a$) fixed points
      for three sample calculations as a function of system size for a
      two-dimensional disordered system.  The number of replicas is
      indicated in parentheses.}
    \label{tab:2D}
\end{table}

\begin{table}[t]
\begin{center}
\begin{tabular}{ |c||c|c|c||c|c|c| } 
 \hline
 Size(No. of replicas) & $W_r(1)$ & $W_r(2)$ & $W_r(3)$ & $W_a(1)$ & $W_a(2)$ & $W_a(3)$  \\ 
 \hline
 $8\times 8 \times 8$($100$) & $5.49$ & $5.46$& $5.23$ & $0.42$ & $0.40$ & $0.49$ \\ 
 $12\times 12 \times 12$($25$) & $4.96$ & $4.87$& $4.73$  & $0.30$ & $0.28$ & $0.25$ \\ 
 $16\times 16 \times 16$($10$) & $4.60$ & $4.96$ & $4.68$ & $0.08$ & $<0.01$ & $0.23$ \\ 
 $20\times 20 \times 20$($5$) & $5.05$ & $4.70$ & $4.58$ & $<0.01$ &  $<0.01$ & $0.05$ \\ 
 $24\times 24 \times 24$($3$) & $4.62$ & $4.97$ & $4.66$ &$<0.01$ &$<0.01$ &$<0.01$ \\ 
 \hline
\end{tabular}
\end{center}
    \caption{The repulsive ($W_r$) and attractive ($W_a$) fixed points
      for three sample calculations as a function of system size for a
      three-dimensional disordered system.  The number of replicas is
      indicated in parentheses.}
    \label{tab:3D}
\end{table}

\section{Conclusions}
\label{sec:cnclsns}

Wegner~\cite{Wegner76} is credited~\cite{Lee85} with introducing
concepts from statistical mechanics into the study of disordered
systems.  In this work we applied statistical mechanical ideas using
the characteristic function of the modern theory of polarization as a
starting point.  In particular we derived a scaling relation according
to the steps followed by Widom to relate critical exponents, and we
applied a renormalization procedure to the problem of disorder.  In 1D
and 3D our method is in full agreement with the common
wisdom~\cite{Abrahams79} on Anderson localized systems, however, in 2D
we encountered system size limitations.

We note that the case of two dimensions has always been the most
difficult one, both
experimentally~\cite{White20,RobertdeSaintVincent10,Jendrzejewski12,Muller15}
and
theoretically~\cite{Kuzovkov02,Markos04,Suslov05,Mott85,Srivastava90}.
Although, it is considered common knowledge that there are no extended
states in two dimensions, the original work of Abrahams {\it et
  al.}~\cite{Abrahams79} states that inspite of the absence of
extended states, due to the crossover between exponential and
logarithmic behavior, experiments may still detect a mobility edge.

\section*{Acknowledgements}

This research was supported by the Ministry of Innovation and
Technology and the National Research, Development and Innovation
Office (NKFIH), within the Quantum Information National Laboratory of
Hungary and the Quantum Technology National Excellence Program
(Project No. 2017-1.2.1-NKP-2017-00001).  BH would like to thank Jan
Wehr for a very enlightening discussion.

\end{document}